\documentclass[aps,pra,preprint,superscriptaddress,longbibliography]{revtex4-2}

\usepackage{amsmath}
\usepackage{amsthm}
\usepackage{amsfonts}
\usepackage{amssymb}
\usepackage[dvipsnames]{xcolor}
\usepackage{graphicx}
\usepackage{tikz}
\usepackage{color}
\usepackage[pdftex]{hyperref} 
\usepackage{longtable}
\usepackage{array}
\usepackage{dsfont}

\begin{document}
	
	\title{Majorana constellations for optical scalar beams and vector fields}

        \author{F. Torres-Leal}
        \thanks{These authors contributed equally to this work.}
        \affiliation{Photonics and Mathematical Optics Group, Tecnologico de Monterrey, Monterrey 64849, Mexico}

        \author{E. Garc\'ia Herrera}
        \thanks{These authors contributed equally to this work.}
        \affiliation{Photonics and Mathematical Optics Group, Tecnologico de Monterrey, Monterrey 64849, Mexico}

        \author{M.~P. Morales Rodr\'iguez}
        \thanks{These authors contributed equally to this work.}
        \affiliation{Departamento de F\'isica Te\'orica, At\'omica y \'Optica, Universidad de Valladolid, 47011 Valladolid, Spain}

        \author{B. Perez-Garcia}
        \email[e-mail: ]{b.pegar@tec.mx}
        \affiliation{Photonics and Mathematical Optics Group, Tecnologico de Monterrey, Monterrey 64849, Mexico} 
        
        \author{B.~M. Rodr\'iguez-Lara}
        \email[e-mail: ]{bmlara@upp.edu.mx; blas.rodriguez@gmail.com}
        \affiliation{Universidad Polit\'ecnica de Pachuca. Carr. Pachuca-Cd. Sahag\'un Km.20, Ex-Hda. Santa B\'arbara. Zempoala, 43830 Hidalgo, Mexico }
	
	\date{\today}

\begin{abstract}
We explore the Majorana stellar representation on the Riemann sphere for  optical scalar beams and vector fields, utilizing the spin and orbital angular momentum of light.
In this framework, star constellations for scalar Laguerre-Gaussian beam basis appear at the poles of the sphere, while those for vector fields correspond to well-defined circular polarizations.
Leveraging the $su(2)$ symmetry of angular momentum, we populate the sphere through generalized unitary rotations of basis elements, producing optical analogues of Bloch and generalized $su(2)$ coherent states for both scalar beams and vector fields.
These rotations position constellations for scalar Hermite-Gaussian beams and vector Hermite-Gaussian partial Poincar\'e fields, restricted to linear polarization, along the equator of the sphere.
We also address the inverse problem, reconstructing optical scalar beams and vector fields from their stellar constellations.
Additionally, we explore scalar and vector analogues of cat codes and kings of quantumness, contributing to the visualization and characterization of complex optical beams and fields. 
These insights suggest potential applications in optical communications.
\end{abstract}

\maketitle
\newpage	

\section{Introduction}

Light propagating through homogeneous, isotropic, and linear media models as a transverse electromagnetic field. 
The electric and magnetic field vectors are perpendicular to each other and to the direction of propagation. 
The temporal behavior of the electric field determines its polarization, characterized by the path traced by the electric field vector over time. 
The most general path is an ellipse in the plane transverse to propagation, known as elliptical polarization. 
This polarization can rotate in either a clockwise or counter-clockwise direction as viewed from the direction of propagation, with special cases including circular and linear polarization \cite{Born2019book, Konstantin2015}.

Using Jones calculus, horizontal and vertical linear polarization vectors correspond to eigenstates of the Pauli-Z matrix. 
The diagonal and anti-diagonal linear polarization vectors correspond with those of the Pauli-X matrix, while the left and right circular polarization vectors align with the eigenstates of the Pauli-Y matrix \cite{Jones1941, Fano1954p121}.
Thus, polarization shows an underlying symmetry provided by the special unitary $su(2)$ Lie algebra with Bargmann parameter $j=1/2$, which matches intrinsic angular momentum \cite{Goldberg2021p1}. 
In this representation, the four Stokes parameters,
\begin{align}
    S_{j} = \langle \hat{\sigma}_{j} \rangle,
\end{align}
are the mean expected value of the identity $\hat{\sigma}_{0}$ and three Pauli matrices $\hat{\sigma}_{j}$ with $j=x,y,z$, and map into the Poincar\'e sphere. 
Each totally polarized state corresponds to a point on the surface of the sphere, whereas those inside correspond to partially polarized states.

The spatial degree of freedom provides another angular momentum component that may be intrinsic or extrinsic \cite{ONeil2002}.
A standard choice to construct a basis for intrinsic orbital angular momentum (OAM) are Laguerre-Gaussian beams (LGBs) \cite{Siegman1986lasers},
\begin{align}
    \begin{aligned}
        \Phi_{p, \ell}(r,\phi,z) =&~ \frac{w_{0}}{w(z)} e^{\frac{-i k r^{2}}{2 R(z)}} e^{ i (2 p + \vert \ell \vert + 1) \varphi(z)} e^{ -\frac{r^{2}}{w^{2}(z)}} \left( \frac{\sqrt{2} r}{w(z)} \right)^{\vert \ell \vert} \mathrm{L}_{p}^{\vert \ell \vert}\left( \frac{2 r^{2}}{w^{2}(z)}\right) e^{i \ell \phi},
    \end{aligned}
\end{align}
for their well-defined hyperbolic \cite{Plick2015} and angular \cite{Allen1992} momenta proportional to the radial number $p$ and the azimuthal number $\ell$, respectively. 
We use the standard definitions for the beam waist $w(z)= w_{0} \sqrt{ 1 + (z/z_{R})^{2}}$, curvature radius $R(z) = z \left[ 1 + (z_{R}/z)^{2} \right]$, and Gouy phase $\varphi(z) = \arctan (z/z_{R})$, in terms of the propagation distance $z$, the propagation number $k = 2 \pi/ \lambda$, the initial waist $w_{0} = \lambda / (\pi \theta)$, and the Rayleigh range $z_{R} = k w_{0}^{2} / 2$, in terms of the wavelength $\lambda$ and beam divergence $\theta$ in the media.

Just as the spin representation of the $su(2)$ Lie algebra describes light polarization, the OAM representation with Dicke states $\vert j; m \rangle$, with Bargmann parameter $j = (2 p + \vert \ell \vert)/2 = 0, 1/2, 1, 3/2,\ldots$ and its projection along the propagation axis $m = \ell / 2 = -j , -j+1,\ldots, j-1, j$, helps our understanding of the orbital angular momentum of light \cite{Arecchi1972}. 
For instance, using Clebsch-Gordan coefficients to calculate the total angular momentum and characterize vector fields that have both polarization and orbital angular momentum. 
However, visualization of these states presents its own challenges. The subspace with Bargmann number $j=0$ is a single point representing the standard Gaussian beam $\psi_{0,0}$. 
The LGBs $\psi_{0,1}$ and $\psi_{0,-1}$, which map into the south and north poles of the Bloch-Poincar\'e sphere, span the subspace with $j=1/2$. 
Subspaces for higher-dimensional orbital angular momenta with Bargmann parameter $j$ require $2j+1$ LGBs as an orthogonal basis, which maps onto a ($2j+1$)-dimensional Bloch hyper-sphere.

Majorana stellar representation provides a geometric representation for states within systems possessing a symmetry with an underlying $su(2)$ Lie algebra \cite{Majorana1931p43,Bloch1945p237, Bengtsson2006}. 
This approach maps a pure state, built through the coherent superposition of OAM Dicke states, into a complex polynomial. 
Projecting the complex roots of this polynomial stereographically onto the Riemann sphere offers a visual representation of the state. 
Thus, a state in a subspace with Bargmann parameter $j$ appears as a constellation of up to $2j$ Majorana stars. 
The Majorana representation has proven invaluable for various applications in quantum physics and quantum optics. 
It has been used to characterize the quantum polarization of light \cite{Bjork2015p031801, Bjork2015p108008}, map structured-Gaussian beams into a modal Majorana constellation \cite{GutierrezCuevas2020p123903}, study the geometrical phases of coherent states of higher-spin \cite{Kam2021p073020}, represent anticoherent spin states \cite{Zimba2006p143}, multipolar highly-quantum states \cite{Romero2024p012214}, and in rotation sensing using king of quantumness states \cite{Kolenderski2008p052333,Bouchard2017p1429, Baecklund2014}. Thus, the Majorana representation is an alternative way to describe quantum states, especially in spin systems.

Visualization techniques for structured light with angular momentum have advanced significantly \cite{He2022}, beginning with the Modal Poincar\'e Sphere, which was developed to represent first-order scalar beams with orbital angular momentum \cite{Padgett1999} and later expanded to third-order modes \cite{Calvo2005}. 
This approach was instrumental in visualizing astigmatic transformations by effectively mapping them onto the sphere \cite{Habraken2010}. 
The Higher-order Poincar\'e Sphere (HOPS) was introduced to address polarization states, offering an extended framework to represent beams with both spin and orbital angular momenta, such as radially and azimuthally polarized cylindrical vector beams, as they propagate through anisotropic or nonlinear media \cite{Milione2011}. 
The Hybrid-order Poincar\'e Sphere (HyOPS) further generalized this concept to accommodate cases where spin and orbital angular momenta interact dynamically in inhomogeneous anisotropic media, allowing for conversion between them \cite{Yi2015}.
Recently, the $SU(2)$ Poincar\'e Sphere was introduced to generalize these models, capturing a broader family of structured light with multidimensional ray-wave structures. This approach reveals a unified view where previous Poincar\'e sphere models are special cases, and it supports higher-dimensional state tailoring for applications across optics and quantum information \cite{Shen2020}.
In contrast, the Majorana stellar representation offers a flexible alternative, allowing any structured Gaussian beam to be visualized as a constellation on the Riemann sphere \cite{GutierrezCuevas2020p123903}.

Here, we revisit Majorana stellar representation to describe optical scalar beams, extend it for vector fields by incorporating total angular momentum, and introduce the inverse problem of reconstructing optical scalar beams and vector fields from their constellations.
First, we show how to construct Majorana constellations for scalar light beams, using LGBs as the building basis, in Section \ref{sec:S2}. 
Then, we populate the Riemann sphere using the optical analogy of Bloch and Generalized Coherent States for the $su(2)$ Lie algebra \cite{MoralesRodriguez2024p1498, MoralesRodriguez2024}, exploring optical analogies for cat codes.
In addition, we introduce the inverse problem of reconstructing optical scalar beams from a Majorana constellation given by the vertices of platonic solids leading to the optical analogues of the so-called king of quantumness states.
Using Clebsch-Gordan coefficients to add the polarization and orbital angular momentum of light into total angular momentum, we build the constellations of full vector fields in Section \ref{sec:S3}.
For the sake of completeness, we explore the vector field analogies of the scalar beams discussed in Section \ref{sec:S2}. 
We close with our conclusions in Section \ref{sec:S5}.

\section{Majorana Constellations for scalar light beams} \label{sec:S2}

Let us begin by discussing the construction of Majorana constellations in the scalar Laguerre-Gaussian beam (LGB) basis. 
We define the orthonormal scalar LGB,
\begin{align}
    \Phi_{p, \ell}(r,\phi,z) = \frac{\sqrt{2}}{w(z)} e^{-i \frac{k r^{2}}{2 R(z)}} e^{i (2p + |\ell| + 1) \varphi(z)} \psi_{n_{+},n_{-}}\left(\frac{\sqrt{2} r}{w(z)}, \phi\right),
\end{align}
in terms of the Laguerre-Gaussian modes (LGMs) \cite{Schwinger2001, MoralesRodriguez2024},
\begin{align}
    \psi_{n_{+},n_{-}}(q_{\rho}, q_{\phi}) = (-1)^{p} \sqrt{\frac{p!}{\pi (p + |\ell|)!}} q_{\rho}^{|\ell|} e^{-\frac{1}{2} q_{\rho}^2} \mathrm{L}_{p}^{|\ell|}(q_{\rho}^2) e^{i \ell q_{\phi}},
\end{align}
which are eigenmodes of the isotropic two-dimensional quantum oscillator  with the left- and right-handed numbers $n_{\pm} = 0, 1, 2, \ldots$. 
We use the shorthand notation $q_{\rho} = \sqrt{2} r /w(z)$ and $q_{\phi} = \phi$ for the sake of space.
The radial and azimuthal numbers reduce to the following expressions,
\begin{align}
    \begin{aligned}
        p =&~ \frac{1}{2} \left( n_{+} + n_{-} - \vert n_{+} - n_{-} \vert \right), \\
        \ell =&~ n_{+} - n_{-}.
    \end{aligned}
\end{align}
The LGMs provide a complete, orthonormal basis,
\begin{align}
    \int_{0}^{\infty} dq_{\rho} \int_{0}^{2\pi} dq_{\phi} ~ q_{\rho} \psi^{\ast}_{m_{+},m_{-}}(q_{\rho}, q_{\phi}) \psi_{n_{+},n_{-}}(q_{\rho}, q_{\phi}) = \delta_{m_{+},n_{+}} \delta_{m_{-},n_{-}},
\end{align}
for the Hilbert space of square-integrable functions on the dimensionless real plane $(q_{\rho}, q_{\phi})$ in polar coordinates. 
At $z=0$, we recover LGMs from LGBs. 
Our analysis focuses on the former, although it remains valid for the latter as well.

It is straightforward to connect LGMs with the OAM Dicke basis \cite{MoralesRodriguez2024p1498,MoralesRodriguez2024},
\begin{align}
    \langle q_{\rho}, q_{\phi} \vert j; m \rangle = \psi_{j + m, j - m}(q_{\rho}, q_{\phi}),
\end{align}
with the parameter $m = -j, -j+1, \ldots, j-1, j$. 
Thus, each subspace comprises $2j+1$ LGMs that conserve the Bargmann parameter $j = (n_{+} + n_{-})/2 = (2 p + \vert \ell \vert)/2$. 
Majorana's stellar representation for visualization of higher-dimensional angular momentum states associates a polynomial \cite{Majorana1931p43},
\begin{align}
    p_{\psi}(\zeta) = \sum_{m=-j}^j (-1)^{j-m} \sqrt{\binom{2j}{j-m}} c_m \zeta^{j+m},
\end{align}
to any state in these Hilbert subspaces,
\begin{align}
    \psi(q_{\rho}, q_{\phi}) = \sum_{m=-j}^j c_m \psi_{j + m, j - m}(q_{\rho}, q_{\phi}),
\end{align}
with $c_{m} \in \mathbb{C}$ such that $\sum_{m=-j}^{j} \vert c_{m} \vert^{2} =1$. 
The Majorana polynomial has $2j$ complex roots $\zeta_j$ as long as $c_m \neq 0$. 
If one or more coefficients $c_m$ are zero, then $\zeta_j=\infty$ is a root of multiplicity equal to the number of zeros. 
Each complex root $\zeta_j$ can be stereographically projected onto the Riemann sphere,
\begin{align}
    \left( x_{j}, y_{j}, z_{j} \right) = \left(\frac{2 \mathrm{Re}(\zeta_{j})}{ 1 + \vert \zeta_{j} \vert^{2} }, \frac{2 \mathrm{Im}(\zeta_{j})}{ 1 + \vert \zeta_{j} \vert^{2} }, \frac{1 - \vert \zeta_{j} \vert^{2} }{1 + \vert \zeta_{j} \vert^{2} }\right),
\end{align}
with roots of zero value mapping onto the north pole and those with infinite value onto the south pole.

\begin{figure}[ht]
\includegraphics[scale=1]{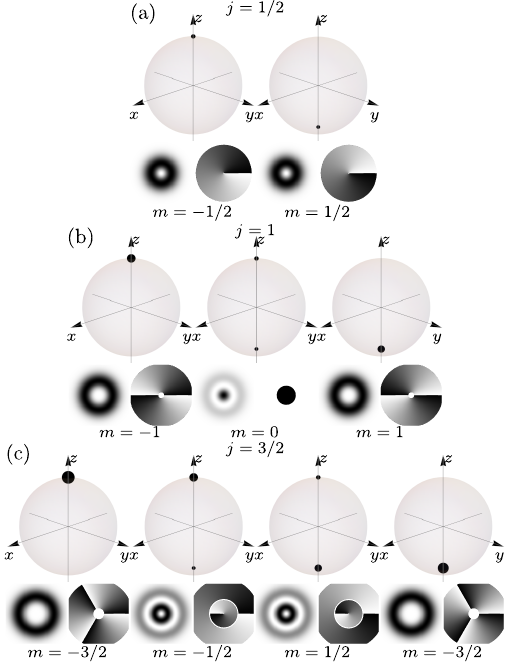}
\caption{Majorana constellation for optical scalar light beam analogue of the OAM Dicke states $\langle q_{\rho}, q_{\phi} \vert j; m \rangle = \psi_{j + m, j - m}(q_{\rho}, q_{\phi})$ spanning the subspace with Bargmann parameter (a) $j = 1/2$ with one star, (b) $j = 1$ with two stars, and (c) $j = 3/2$ with three stars per constellation, as well as the intensity and phase distribution for the corresponding LGM. The size of the point on the Riemann sphere is directly proportional to the number of stars.} \label{fig:Figure1}
\end{figure}

We slightly deviate from Majorana's method. 
Usually, any given state in the Dicke basis $\vert j; m \rangle$, corresponding to the LGM with $n_{+}= j + m$ and $n_{-}= j - m$, maps onto $j-m$ stars at the north pole with no stars at the south pole, except for $m = j$ which maps into $2j$ stars at the south pole with no stars at the north pole. 
Instead, we introduce $j+m$ stars in the south pole to create constellations composed of $2j$ Majorana stars. 
The state $\vert j; m \rangle$ is represented by $j-m$ stars at the north pole with $j+m$ stars at the south pole. 
This arrangement allows us to use the number of stars at the poles to determine the left- and right-handed numbers of the LGM, facilitating the recovery of the radial and azimuthal numbers. 
Figure \ref{fig:Figure1} illustrates the Majorana constellations for the Dicke states $\vert j, m \rangle$ spanning the subspaces with Bargmann parameter $j=1/2, 1, 3/2$. 
The size of the point on the Riemann sphere is directly proportional to the number of stars. 
We also show the intensity and phase distribution for the corresponding LGMs.

\subsection{Bloch and Generalized Coherent States} 

Now, let us populate the Riemann sphere. 
The $su(2)$ Lie algebra provides a general unitary rotation \cite{VillanuevaVergara2015, MoralesRodriguez2024}, 
\begin{align}
    \begin{aligned}
        \hat{R}(\vartheta, \varphi) =&~ e^{\vartheta \left( e^{i \varphi}  \hat{J}_{+} - e^{-i \varphi}  \hat{J}_{-} \right)}, \\
        =&~ e^{ i \varphi \hat{J}_{z}} e^{ i 2 \vartheta \hat{J}_{y}} e^{-i \varphi \hat{J}_{z}}, \\
        =&~ \sum_{p,q=0}^{2j} (-1)^{(j-p)} e^{i \varphi (p-q)} \sin^{2j} \vartheta \tan^{-(p+q)} \vartheta \sqrt{ \binom{2j}{j-p} \binom{2j}{j-q}} \times \ldots \\ 
        &~\ldots \times {}_{2}\mathrm{F}_{1}\left[ p-j, q-j; -2j; \csc^{2} \vartheta \right] \vert j; p \rangle \langle j; q \vert,
    \end{aligned}
\end{align}
which corresponds to a $2 \vartheta$ rotation around the $y$-axis and a $\varphi$ rotation around the $z$-axis of the sphere. 
In this context, $\vartheta$ serves as the polar angle, ranging from $0$ to $\pi/2$, and $\varphi$ as the azimuthal angle, ranging from $0$ to $2\pi$.

\begin{figure}
\includegraphics[scale=1]{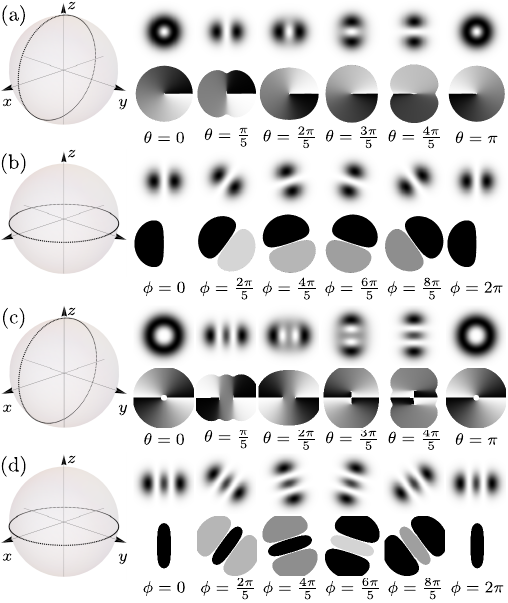}
\caption{Majorana constellations, irradiance, and phase distributions for the optical scalar light beam analog of Bloch coherent states (a) and (c) $\psi_{\vartheta,0} (q_{\rho}, q_{\phi})$ and (b) and (d) $\psi_{\pi/4,\varphi} (q_{\rho}, q_{\phi})$ for (a)-(b) $j=1/2$ and (c)-(d) $j=1$.} \label{fig:Figure2}
\end{figure}

This rotation acting on the lowest LGMs basis of the corresponding OAM Dicke basis, 
\begin{align}
    \begin{aligned}
        \psi_{\vartheta,\varphi} (q_{\rho}, q_{\phi})
        =&~ \left( e^{i \varphi} \sin \vartheta \right)^{2j} \sum_{k=0}^{2j} \sqrt{\left( \begin{array}{c} 2j \\ k \end{array} \right)} \left( e^{i \varphi} \tan \vartheta \right)^{-k}   \psi_{2j-k,k} (q_{\rho}, q_{\phi}),
    \end{aligned}
\end{align}
yields the optical analogy of Bloch coherent states \cite{MoralesRodriguez2024p1498}. 
They form an over complete,
\begin{align}
    \begin{aligned}
        \int_{0}^{\infty} dq_{\rho} \, q_{\rho} \int_{0}^{2 \pi} dq_{\phi} \, \psi_{\tilde{\vartheta},\tilde{\varphi}}^{\ast} (q_{\rho}, q_{\phi}) \psi_{\vartheta,\varphi} (q_{\rho}, q_{\phi}) =&~ \left( \csc \vartheta \csc \tilde{\vartheta} \right)^{j} \\
        &~\left[ e^{\frac{i}{2}(\varphi - \tilde{\varphi})} \sin \vartheta \sin \tilde{\vartheta} + e^{-\frac{i}{2}(\varphi - \tilde{\varphi})} \cos \vartheta \cos \tilde{\vartheta} \right]^{2j},
    \end{aligned}
\end{align}
basis for states for the Hilbert space subspace with Bargmann parameter $j$.
Figure \ref{fig:Figure2} shows optical analogy of Bloch coherent states $\psi_{\vartheta,0} (q_{\rho}, q_{\phi})$ and $\psi_{\pi/4,\varphi} (q_{\rho}, q_{\phi})$ for Bargmann parameters $j=1/2, 1$. 

Generalized Gilmore-Perelomov coherent states for the $su(2)$ Lie algebra arise from the rotation of any LGMs of the OAM Dicke basis \cite{VillanuevaVergara2015}, 
\begin{align}
    \begin{aligned}
        \psi_{\vartheta,\varphi, m} (q_{\rho}, q_{\phi}) =&~ \sqrt{\left( \begin{array}{c} 2j \\ j-m \end{array} \right)} \, \sec^{-2 j} \vartheta  \left( e^{ i \varphi} \tan \vartheta \right)^{j} \times \\
        &\times \sum_{k=0}^{2j} (-1)^{k} \sqrt{ \left( \begin{array}{c} 2j \\ k \end{array} \right)  } \,  e^{ - i \varphi (k+m)} \tan^{(k-m)} \vartheta   \,  \times \\
        & \times  \,_{2}\mathrm{F}_{1} \left(-k,-j+m; -2j; \csc^{2} \vartheta \right) \, \psi_{2j-k,k} (q_{\rho}, q_{\phi}), 
    \end{aligned}
\end{align}
that allow us to populate the Riemann sphere with less clustered constellations. 
For example, if we rotate the LGM $\psi_{k,2j-k} (q_{\rho}, q_{\phi})$, that has a constellation with $k$ stars in the south pole and $2j-k$ stars in the north pole, we obtain a constellation with all its stars rotated by a $\vartheta$ angle around the $z$-axis and $\varphi$ angle around the $y$-axis.

It is straightforward to realize that the Hermite-Gauss modes (HGMs) are generalized coherent states of the LGMs \cite{Abramochkin2004}, 
\begin{align}
    \begin{aligned}
        \psi_{n_{x},n_{y}}^{(HG)}(q_{x}, q_{y}) =&~ (- i)^{n_{x}} \psi_{\pi/4,0, (n_{x}-n_{y}/2)} (q_{\rho}, q_{\phi}),\\
        =&~ (- i)^{n_{x}} 2^{-\frac{1}{2} (n_{x} + n_{y})} \sqrt{ \binom{n_{x} + n_{y}}{n_{y}} }  \times \\
        &~ \times \sum_{k=0}^{2j} (-1)^{k} \sqrt{\binom{n_{x} + n_{y}}{k}} {}_{2}\mathrm{F}_{1}\left[ -n_{y}, -k; -n_{x}-n_{y}; 2  \right] \psi_{n_{x}+n_{y}-k,k} (q_{\rho}, q_{\phi}),
    \end{aligned}
\end{align}
in the Bargmann subspace $j= (n_{x} + n_{y})/2$ with horizontal $n_{x} = j - m$ and vertical $n_{y} = j + m$ numbers.
Thus, the HGMs Majorana constellation will consist of $n_{x}$ stars at the point $(x,y,z)=(1,0,0)$ and $n_{y}$ stars at the point $(-1,0,0)$ on the Riemann sphere.
The equator of the Riemann sphere will contain  rotated HGMs with inclination angle equal to the azimuthal angle $\varphi$, Fig. \ref{fig:Figure2}(b) and Fig. \ref{fig:Figure2}(d).

\begin{figure}
\includegraphics[scale=1]{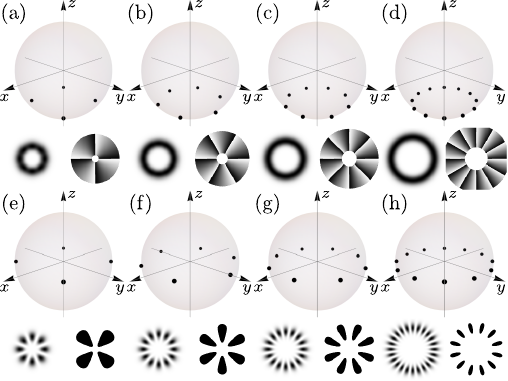}
\caption{Majorana constellations, irradiance, and phase distributions for scalar light beams analogues of the zeroth cat code with (a)-(d) $\vartheta = \pi/8$ and (e)-(h) $\vartheta = \pi/4$, and $\varphi=0$ for (a),(e)  $j=2$, (b),(f)  $j=3$, (c),(g)  $j=4$, and (d),(h)  $j=6$.} \label{fig:Figure3}
\end{figure}

\subsection{Cat codes and Kings of Quantumness} 

In quantum information theory, cat codes refer to quantum error correction codes that utilize superposition of coherent states \cite{Li2017p030502,Leghtas2013p120501,Mirrahimi2014p045014,Grimsmo2020p011058},
\begin{align}
    \begin{aligned}
        \vert C_{\alpha}^{n} \rangle = \frac{1}{2 d \sqrt{\mathcal{N}_{\alpha}^{n}}} \sum_{k=0}^{2d-1} e^{-\frac{i \pi}{d} k n} \vert e^{\frac{i \pi}{d} k} \alpha \rangle,
    \end{aligned}
\end{align}
where $\alpha$ is the coherent parameter, $\mathcal{N}$ is a normalization constant, the size of the basis is $2d$ with $d\ge 2$, and $n = 0, 1, 2, \ldots, 2d-1$.
They are an extension to Schr\"odinger cat states, designed to be robust against phase errors.
It is straightforward to recognize that the optical analogy of these codes in the Hilbert subspaces with constant Bargmann parameter $j$,
\begin{align}
    \begin{aligned}
        \vert C_{\vartheta, \varphi}^{m} \rangle = \frac{1}{\sqrt{\mathcal{N}_{\vartheta,\varphi}^{m}}} \sum_{k=0}^{2j-1} e^{-\frac{i \pi}{j} k m} \hat{R}(\vartheta, \varphi + \pi k / j) \vert j; -j \rangle,
    \end{aligned}
\end{align}
with sum modulus $2 \pi$ for the phase,  normalization constant $\mathcal{N}_{\vartheta,\varphi}^{m}$, and cat code index $m = 1, 2, \ldots, 2j-1$, yield the LGMs spanning the subspace,
\begin{align}
    \begin{aligned}
        \langle q_{\rho}, q_{\phi} \vert C_{\vartheta, \varphi}^{0} \rangle =&~ \frac{1}{\sqrt{\cos^{4j} \vartheta + \sin^{4j} \vartheta}} \left[  \cos^{2j} \vartheta ~ \psi_{2j , 0}(q_{\rho}, q_{\phi}) + \left( e^{i \varphi} \sin \vartheta \right)^{2j} ~ \psi_{0, 2j}(q_{\rho}, q_{\phi}) \right],\\
    \langle q_{\rho}, q_{\phi} \vert C_{\vartheta, \varphi}^{m} \rangle =&~ e^{i m \varphi} \psi_{2j-m,m} \left( q_{\rho}, q_{\phi}\right),
    \end{aligned}
\end{align}
with the exception of the zeroth cat code, which provides a superposition of the first and last OAM Dicke states in the basis. 
The Majorana constellation for the zeroth cat code appears on the $2 (\pi/2-\vartheta)$ parallel of the Riemann sphere, with its stars distributed equidistantly and rotated by an azimuthal angle $\varphi$, Fig. \ref{fig:Figure3}.

\begin{figure}
\includegraphics[scale=1]{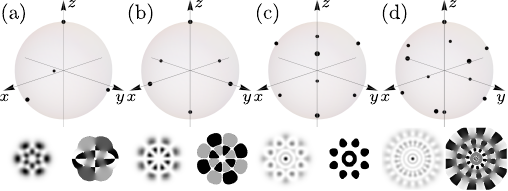}
\caption{Majorana constellations, irradiance, and phase distributions for scalar light beams which stars locate at the vertices of four platonic solids: (a) tetrahedron $j=2$, (b) octahedron $j=3$, (c) hexahedron $j=4$, and (d) dodecahedron $j=6$.} \label{fig:Figure4}
\end{figure}

It is also possible to produce a scalar light beam from a given Majorana constellation, effectively solving an inverse problem. 
First, we recover the roots $\zeta_{j}$ of the Majorana polynomial from the Majorana constellation by applying the stereographic projection,
\begin{align}
    \zeta_{j} = \frac{x_{j} + i y_{j}}{1 + z_{j}},
\end{align}
that maps points $(x_{j},y_{j},z_{j})$ on the Riemann sphere to the complex plane, where the north pole $(0,0,1)$ maps to the origin and the south pole $(0,0,-1)$ to a point at infinity.
These complex values serve as the roots of the Majorana polynomial, 
\begin{align}
    \begin{aligned}
        p_{\psi}(\zeta) =&~ d_{j} \prod_{m=-j}^{j} (\zeta - \zeta_{m}), \\
        =&~ \sum_{m=-j}^j d_{m} \zeta^{j+m},
    \end{aligned}
\end{align}
associated with the scalar light beam $\psi$, where we define the shorthand notation $d_{m} = (-1)^{j-m} \sqrt{\binom{2j}{j-m}} c_{m}$.
Using, Vieta's formulas \cite{Funkhouser1930},
\begin{align}
\frac{d_{m}}{d_{j}} = (-1)^{j-m} \sum_{-j \le p_{-j} \le p_{-j+1} \le \ldots \le j } \prod_{q=-j}^{j} \zeta_{p_{q}+j+1}, 
\end{align}
we can determine the modified coefficients $d_{m}$ for $m = -j, -j+1, \ldots j-1$, while we recover the final coefficient $d_{j}$ from the normalization condition $\sum_{m=-j}^{j} \vert d_{m} \vert^{2} = 1$.
From the modified coefficients $d_{m}$, it is straightforward to recover recover the coefficients $c_{m}$,
\begin{align}
c_{m}= (-1)^{m} \sqrt{\frac{j! (j+m)!}{(2j)!}} d_{m}, 
\end{align}
for the superposition of the corresponding LGMs in the Bargmann subspace.
For example, starting from a Majorana constellation given by the vertices coordinates of a Platonic solid in the Riemann sphere \cite{GutierrezCuevas2020}, we can generate their corresponding scalar light beam, Fig. \ref{fig:Figure4}.
These states are the optical analogues of the kings of quantumness used for sensing rotations while establishing the coordinate system for quantum communications \cite{Kolenderski2008p052333,Bouchard2017p1429}.

\section{Majorana Constellations for Vector Fields} \label{sec:S3}

It is a standard occurrence in polarimetry to map polarization onto the Poincar\'e sphere. 
This is equivalent to assigning polarization an intrinsic angular momentum representation with Bargmann parameter $j=1/2$, such that the states $\vert 1/2, \pm 1/2 \rangle \equiv \boldsymbol{\epsilon}_{\pm 1/2}$ represent left- and right-handed circular polarization. 
Integrating the polarization and spatial degrees of freedom leads to light fields,
\begin{align}
    \begin{aligned}
        \boldsymbol{\psi}\left( q_{\rho}, q_{\phi}\right) = \sum_{j=0}^{\infty} \sum_{m=-j}^{j} \sum_{k=-1/2}^{1/2} c_{j,m,k} \psi_{j+m,j-m} \left( q_{\rho}, q_{\phi}\right) \boldsymbol{\epsilon}_{k},
    \end{aligned}
\end{align}
constructed, for example, by the superposition of spatially dependent scalar Laguerre-Gaussian beams and spatially independent polarization vectors.
The analogy of polarization and spatial degrees of freedom with spin and angular momenta allows us to rewrite these light fields,
\begin{align}
    \begin{aligned}
       \boldsymbol{\psi}\left( q_{\rho}, q_{\phi}\right) = \sum_{J=0}^{\infty} \sum_{M=-J}^{J} C_{J;M} \boldsymbol{\Psi}_{J;M}\left( q_{\rho}, q_{\phi}\right),
    \end{aligned} \label{eq:LightField}
\end{align}
in the total angular momentum (TAM) Dicke basis,
\begin{align}
    \boldsymbol{\Psi}_{J;M}\left( q_{\rho}, q_{\phi}\right) = \sum_{m_{s} = -j_{s}}^{j_{s}} \sum_{m_{p} = -1/2}^{1/2} c_{j_{s},m_{s},j_{p},m_{p}}^{J,M}  \psi_{j_{s}+m_{s},j_{s}-m_{s}} \left( q_{\rho}, q_{\phi}\right) \boldsymbol{\epsilon}_{m_{p}}, \label{eq:TAMState}
\end{align}
where the Clebsch-Gordan coefficients $c_{j_{s},m_{s},j_{p},m_{p}}^{J,M}$ provide the precise weights for each possible combination of spin and orbital angular momenta required to construct a well-defined total angular momentum state \cite{Alex2011arXiv, Griffiths2019}.

\begin{figure}[ht]
\includegraphics[scale=1]{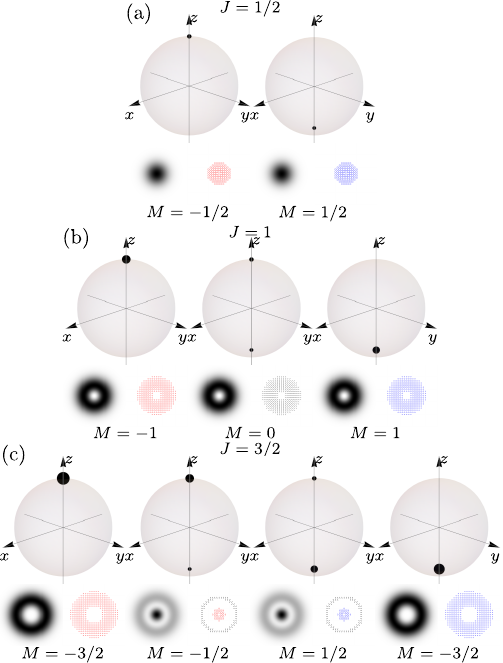}
\caption{(Color online) Majorana constellations for vector light fields $\boldsymbol{\Psi}_{J,M}\left( q_{\rho}, q_{\phi}\right)$ analogues to TAM Dicke states, spanning subspace with Bargmann parameters (a) $J = 1/2$ with one star, (b) $J = 1$ with two stars, and (c) $J = 3/2$ with three stars per constellation.
Each set shows the Majorana constellation (top), alongside the intensity (bottom left) and polarization distribution (bottom right) of the vector light field.
Red, blue and black indicates left-handed, right-handed, and linear polarization, respectively.}
\label{fig:Figure5}
\end{figure}

\begin{figure}
    \centering
    \includegraphics[scale=1]{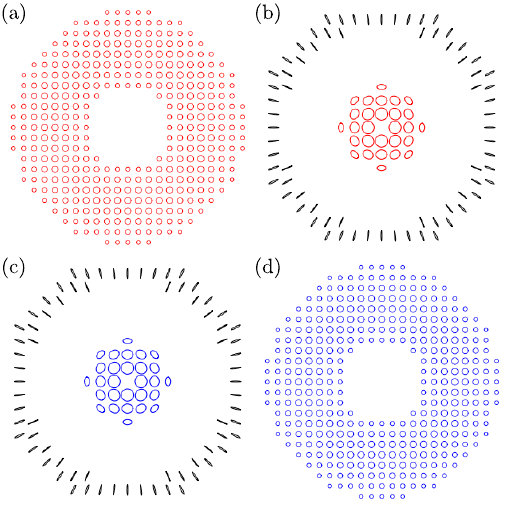}
    \caption{(Color online) Polarization distribution for the vector light fields $\boldsymbol{\Psi}_{J,M}\left( q_{\rho}, q_{\phi}\right)$ analogues to TAM Dicke states with parameters $\{J,M\}$ equal to (a) $\{3/2, -3/2\}$, (b) $\{ 3/2, -1/2\}$, (c) $\{3/2, 1/2\}$, and (d) $ \{3/2, -3/2\}$. Red, blue and black indicates left-handed, right-handed, and linear polarization, respectively.}
    \label{fig:Figure6}
\end{figure}

A TAM Bargmann parameter $J$ decomposes into polarization and spatial parameters $j_{p} = 1/2$ and $j_{s}= \vert J - 1/2 \vert$, respectively. 
Thus, TAM $J=0$ corresponds to the singlet $\left[ \psi_{1,0} \left( q_{\rho}, q_{\phi}\right) \boldsymbol{\epsilon}_{-1/2} - \psi_{0,1} \left( q_{\rho}, q_{\phi}\right) \boldsymbol{\epsilon}_{1/2} \right]/2$ providing linear polarization for $M=0$.
TAM $J=1/2$ corresponds to a spatial $j_{s} = 0$; that is, the Gaussian beam $\psi_{0,0} \left( q_{\rho}, q_{\phi}\right) \boldsymbol{\epsilon}_{\pm1/2}$ providing right- and left-handed circular polarization for $M=-1/2$ and $M=1/2$, respectively, Fig. \ref{fig:Figure5}(a). 
TAM $J=1$ corresponds to a spatial $j_{s} = 1/2$ with $m_{s}=-1/2$, yielding $\psi_{0,1} \left( q_{\rho}, q_{\phi}\right) \boldsymbol{\epsilon}_{-1/2}$, providing right-handed circular polarization for $M=-1$, the superposition of $m_{s}=\pm 1/2$ with right- and left-handed circular polarization $\left[ \psi_{1,0} \left( q_{\rho}, q_{\phi}\right) \boldsymbol{\epsilon}_{-1/2} + \psi_{0,1} \left( q_{\rho}, q_{\phi}\right) \boldsymbol{\epsilon}_{1/2} \right]/2$, providing linear polarization for $M=0$, and $m_{s}=1/2$, yielding $\psi_{1,0} \left( q_{\rho}, q_{\phi}\right) \boldsymbol{\epsilon}_{-1/2} \boldsymbol{\epsilon}_{1/2}$, providing left-handed circular polarization for $M=1$, Fig. \ref{fig:Figure5}(b). 
The extremes of the TAM ladder, $M=\pm J$, show well-defined right- and left-handed circular polarization, while the intermediate states show a coherent superposition of them, Fig. \ref{fig:Figure5}(c).
For visualization and clarity purposes, Fig.\ \ref{fig:Figure6} presents the polarization distribution for the vector light fields $\boldsymbol{\Psi}_{J,M}\left( q_{\rho}, q_{\phi}\right)$ analogues to TAM Dicke states with parameters $\{J,M\}$, with scaling adjusted for illustrative purposes.

\subsection{Bloch and Generalized Coherent States}

\begin{figure}
\includegraphics[scale=1]{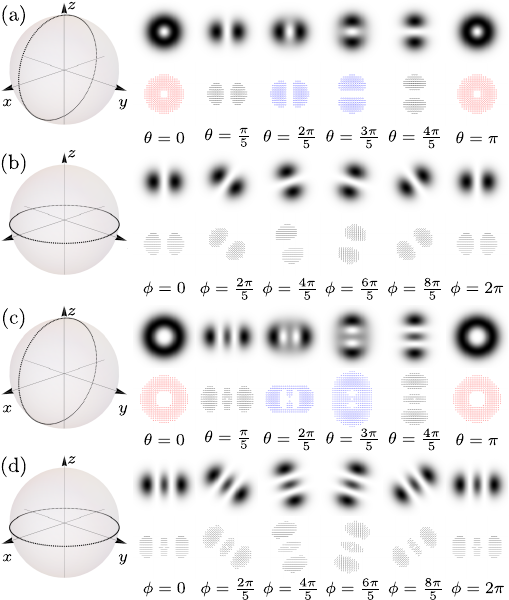}
\caption{(Color online) Majorana constellations for the vector light fields analog of Bloch coherent states (a) and (c) $\boldsymbol{\Psi}_{\vartheta,0} (q_{\rho}, q_{\phi})$ and (b) and (d) $\boldsymbol{\Psi}_{\pi/4,\varphi} (q_{\rho}, q_{\phi})$ for (a)-(b) $j=1$ and (c)-(d) $j=3/2$.
Each set depicts the Majorana constellation (left), intensity (top), and polarization distribution (bottom) for the vector light field.
Red, blue and black indicate left-handed, right-handed, and linear polarization, respectively.}
\label{fig:Figure7}
\end{figure}

To populate the Riemann sphere for vector fields, we use the general unitary rotation for the $su(2)$ algebra $\hat{R}(\vartheta, \varphi)$ acting on the lowest vector field mode $\boldsymbol{\Psi}_{J;-J}\left(q_{\rho}, q_{\phi} \right)$ to obtain,  
\begin{align}
    \begin{aligned}
        \boldsymbol{\Psi}_{\vartheta,\varphi} \left(q_{\rho}, q_{\phi} \right) =&~ \left( e^{i \varphi} \sin \vartheta \right)^{2j} \sum_{k=0}^{2J} \sqrt{\left( \begin{array}{c} 2J \\ k \end{array} \right)} \left( e^{i \varphi} \tan \vartheta \right)^{-k} \boldsymbol{\Psi}_{2J-k;k} (q_{\rho}, q_{\phi}),
    \end{aligned}
\end{align}
vector light fields analogues of Bloch coherent states.
Figure \ref{fig:Figure7} shows vector fields  $\boldsymbol{\Psi}_{\vartheta,0} (q_{\rho}, q_{\phi})$ and $\boldsymbol{\Psi}_{\pi/4,\varphi} (q_{\rho}, q_{\phi})$ for TAM Bargmann parameters $J=1, 3/2$.
Only the poles possess a well defined polarization a generalized coherent state shows a position dependent polarization.
The equator correspond to partial Poincar\'e fields restricted to linear polarization with Hermite-Gaussian spatial profile \cite{Alonso2017}, Fig. \ref{fig:Figure7}(b) and Fig. \ref{fig:Figure7}(d).

Generalized Gilmore-Perelomov coherent states for the $su(2)$ Lie algebra arise from the rotation of any vector field of the TAM Dicke basis \cite{Perelomov1972, Zhang1990},
\begin{align}
    \begin{aligned}
        \boldsymbol{\Psi}_{\vartheta,\varphi, M} (q_{\rho}, q_{\phi}) =&~ \sqrt{\left( \begin{array}{c} 2J \\ J-M \end{array} \right)} \, \sec^{-2 J} \vartheta  \left( e^{ i \varphi} \tan \vartheta \right)^{J} \times \\
        &\times \sum_{k=0}^{2J} (-1)^{k} \sqrt{ \left( \begin{array}{c} 2j \\ k \end{array} \right)  } \,  e^{ - i \varphi (k+M)} \tan^{(k-M)} \vartheta \, \times \\
        & \times  \,_{2}\mathrm{F}_{1} \left(-k,-J+M; -2j; \csc^{2} \vartheta \right) \, \boldsymbol{\Psi}_{2J-k;k} (q_{\rho}, q_{\phi}),
    \end{aligned}
\end{align}
which allow us to populate the Riemann sphere with less clustered constellations. 
For example, if we rotate the LGM $\boldsymbol{\Psi}_{J;-J+k} (q_{\rho}, q_{\phi})$,which has a constellation with $k$ stars in the south pole and $2J-k$ stars in the north pole, the resulting field will have an equivalent antipodal distribution.

\subsection{Cat codes and Kings of Quantumness}

\begin{figure}
\includegraphics[scale=1]{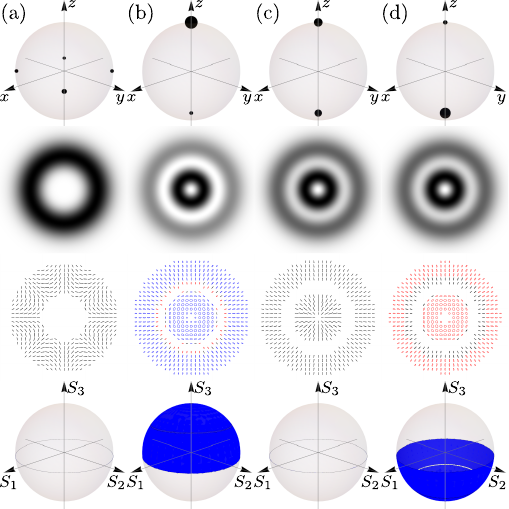}
\caption{(Color online) Majorana constellations for vector light fields analogues to (a) zeroth through (d) third cat codes with $\vartheta = \pi/4$ and $\varphi=0$ for $J=2$.
Each set depicts the Majorana constellation (first row), intensity (second row) and polarization (third row) distributions, and Poincar\'e sphere (fourth row) for the vector light field.
Red, blue and black indicate left-handed, right-handed, and linear polarization, respectively.}
\label{fig:Figure8}
\end{figure}

\begin{figure}
    \centering
    \includegraphics[scale=1]{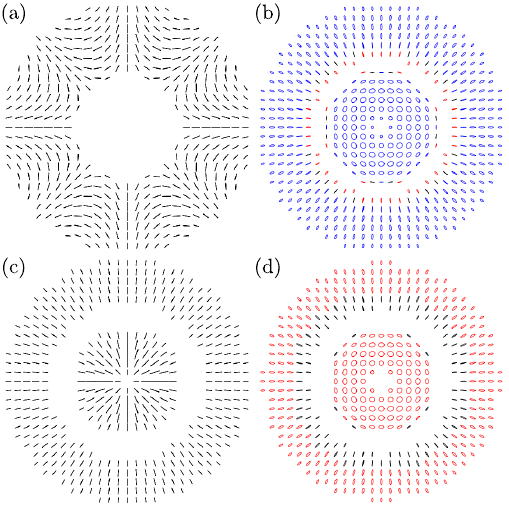}
    \caption{(Color online) Polarization distribution for vector light fields analogues to (a) zeroth through (d) third cat codes with parameters $\vartheta=\pi/4$, $\varphi=0$ for $J=2$. Red, blue and black indicates left-handed, right-handed, and linear polarization, respectively.}
    \label{fig:Figure9}
\end{figure}

As in the scalar case, vector cat codes,
\begin{align}
    \begin{aligned}
        \langle q_{\rho}, q_{\phi} \vert C_{\vartheta, \varphi}^{0} \rangle =&~ \frac{1}{\sqrt{\cos^{4j} \vartheta + \sin^{4j} \vartheta}} \left[ \cos^{2j} \vartheta ~ \boldsymbol{\Psi}_{J;J}(q_{\rho}, q_{\phi}) + \left( e^{i \varphi} \sin \vartheta \right)^{2j} ~ \boldsymbol{\Psi}_{J;-J}(q_{\rho}, q_{\phi}) \right],\\
    \langle q_{\rho}, q_{\phi} \vert C_{\vartheta, \varphi}^{m} \rangle =&~ e^{i m \varphi} \boldsymbol{\Psi}_{J;J-m} \left( q_{\rho}, q_{\phi}\right),
    \end{aligned}
\end{align}
yield the vector light fields analogues of TAM Dicke states, with the exception of the zeroth cat code, which provides a superposition of the first and last TAM Dicke states in the basis. 
Figure \ref{fig:Figure8} shows the four vector field cat codes for Bargmann parameter $J=2$.
The first row shows the Majorana constellation on the Riemann sphere.
The second row shows the irradiance distribution provided by the zeroth component of the stokes parameters $S_{0}$.
The third row shows the polarization distribution in the plane transverse to propagation.
The fourth row shows the polarization states in the Poincar\'e sphere.
None of the vector light fields cat codes have well defined polarization.
The zeroth, Fig. \ref{fig:Figure8}(a), and second, Fig. \ref{fig:Figure8}(c), are partial Poincar\'e fields showing all possible linear polarizations.
The first, Fig. \ref{fig:Figure8}(b), and fourth, Fig. \ref{fig:Figure8}(d), show partial Poincar\'e fields showing all possible left- and right-handed polarizations, in that order.
For visualization and clarity purposes, Fig.\ \ref{fig:Figure9} presents the polarization distribution for the vector light fields analogues of the zeroth to third cat codes with parameters $\vartheta=\pi/4$, $\varphi=0$ for $J=2$, with scaling adjusted for illustrative purposes.

\begin{figure}
\includegraphics[scale=1]{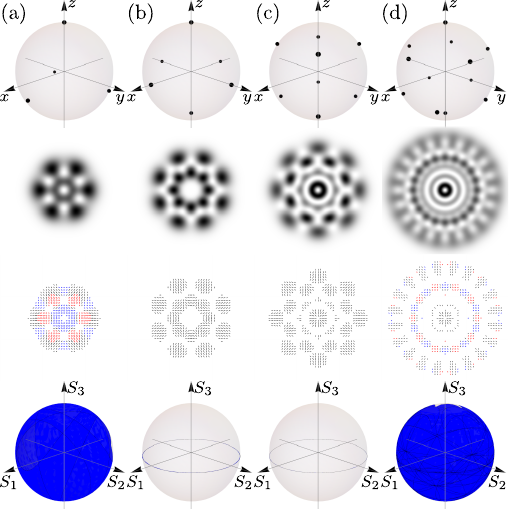}
\caption{(Color online) Majorana constellations for vector light fields with stars located at the vertices of four platonic solids, (a) tetrahedron $j=2$, (b) octahedron $j=3$, (c) hexahedron $j=4$, and (d) dodecahedron $j=6$.
Each set depicts the Majorana constellation (first row), intensity (second row) and polarization (third row) distributions, and Poincar\'e sphere (fourth row) for the vector light field.
Red, blue and black indicate left-handed, right-handed, and linear polarization, respectively.}
\label{fig:Figure10}
\end{figure}

\begin{figure}
    \includegraphics[scale=1]{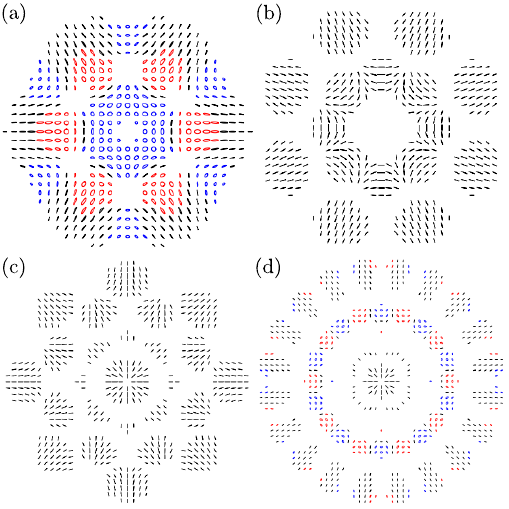}
    \caption{Polarization distribution for vector light fields with stars located at the vertices of four platonic solids, (a) tetrahedron $j=2$, (b) octahedron $j=3$, (c) hexahedron $j=4$, and (d) dodecahedron $j=6$. Red, blue and black indicates left-handed, right-handed, and linear polarization, respectively.}
    \label{fig:Figure11}
\end{figure}

It is also possible to produce vector light fields from a Majorana constellation using the procedure outlined in the previous section.
The solution to the scalar inverse problem provides the coefficients $C_{J,M}$ for the light field, Eq. \ref{eq:LightField}, corresponding to the Majorana constellation. 
Then, we substitute the total angular momentum light field, Eq. \ref{eq:TAMState}, to recover the polarization and spatial modes components of the light field.
For example, starting from the Majorana constellation composed by the vertices of a Platonic solid in the Riemann sphere, we generate the corresponding vector light field \cite{GutierrezCuevas2020}, Fig. \ref{fig:Figure10}.
These states are the optical field analogues of the kings of quantumness and do not show a well defined polarization. 
Those whose Majorana constellation possess reflection symmetry with respect to the three axis in the Riemann sphere show all possible linear polarization, Fig. \ref{fig:Figure10}(b) and Fig. \ref{fig:Figure10}(c).
Those without the symmetry cover most of the Poincar\'e sphere with two missing polarization regions keeping them from becoming full Poincar\'e fields.
For clarity in visualization, Fig.\ \ref{fig:Figure11} shows the polarization of vector light fields with star constellations located at the vertices of four platonic solids, with scaling adjusted for illustrative purposes.

\section{Conclusion} \label{sec:S5}

We used Majorana stellar representation for angular momentum states to construct and visualize optical scalar beams and vector fields as Majorana constellations on the Riemann sphere. 
This approach leverages the analogy between the polarization and spatial degrees of freedom of light with spin and orbital angular momenta, providing a systematic framework to visualize higher-order orbital angular momentum scalar beams and total angular momentum vector fields.

Scalar Laguerre-Gaussian beams provide orthogonal bases for subspaces with constant orbital angular momentum, mapping onto constellations residing at the poles of the Riemann sphere.
Constellations for scalar Hermite-Gaussian beams, on the other hand, reside at the equator. 
Constellations on the sphere are generated through generalized unitary rotations, and their coherent superpositions, provided by the $su(2)$ Lie algebra structure of these subspaces. 
Interestingly, scalar optical analogues of cat codes, quantum states robust to dephasing, reduce to Laguerre-Gaussian beams.
On the other hand, analogues to kings of quantumness, used to align reference frames in quantum communications, exhibit irradiance and phase distributions that allow identifying rotations in the plane transverse to propagation. 

Including the circular polarization basis yields orthogonal bases for subspaces with constant total angular momentum. 
Constellations residing at the poles of the Riemann sphere correspond to vector fields with well-defined circularly polarized light, while those on the equator correspond to partial Poincar\'e fields showing all possible linear polarization states. 
Constellations on the sphere are generated through generalized unitary rotations and their coherent superpositions. 
Vector field analogues of cat codes are partial Poincar\'e fields restricted to linear, left-, or right-handed polarization states. 
Analogues to kings of quantumness show irradiance distributions that allow identifying rotations.
Additionally, those with constellations symmetric with respect to the coordinate axes are partial Poincar\'e fields showing all linear polarization states, while those without such a symmetry display almost all possible polarization states. 

Experimental construction of the optical scalar beams and vector fields represented by our theory can be achieved through established techniques in optics. 
Scalar beams require complex amplitude modulation, which can be implemented using phase-only Spatial Light Modulators (SLMs) or Digital Micromirror Devices (DMDs). 
Computer-generated holography, a widely available method in optics laboratories, provides an effective means for this purpose \cite{Arrizon2007, Forbes2016, GutierrezCuevas2024}.
Vector fields require precise control over both the transverse complex amplitude and electromagnetic field orientation. 
A standard approach involves preparing each orthogonal component of the electromagnetic field separately and then combining them through an interferometric setup \cite{Tidwell1990}. 
Recent advancements have introduced additional methods for this process \cite{Maurer2007, Chen2015, Moreno2015, Rosales2020}. 
Finally, spatially varying polarization states across the transverse plane can be measured using conventional Stokes polarimetry \cite{Goldstein2017, Singh2020}.

We believe Majorana stellar representation offers a powerful framework to visualize, analyze, and propose complex optical beams and fields with higher-order angular momentum, creating a bridge between optical physics and quantum optics to enhance our understanding and control in the tailoring of light.

\section*{Funding}
Not applicable.

\section*{Acknowledgment}
B.~M.~R.~L. is grateful to Miguel \'Angel Avenda\~no Bernal for fruitful discussion. 

\section*{Disclosures} 
The authors declare no conflicts of interest.

\section*{Data availability} 
Data underlying the results presented in this paper may be obtained from the authors upon reasonable request.


%

\end{document}